# Direct Imaging of Resonant Phonon-Magnon Coupling

Chenbo Zhao[1,2], Zhizhi Zhang[1], Wei Zhang[3,1], Yi Li[1], John E. Pearson[1], Ralu Divan[4], Jianbo Wang[2], Valentine Novosad[1], Qingfang Liu[2†], and Axel Hoffmann[1,5‡][1]

[1] Materials Science Division, Argonne National Laboratory, Argonne, Illinois 60439, USA

[2] Key Laboratory of Magnetism and Magnetic Materials of the Ministry of Education, Lanzhou University, Lanzhou 730000, People's Republic of China

[3] Department of Physics, Oakland University, Rochester, MI 48309, USA

[4] Center for Nanoscale Materials, Argonne National Laboratory, Argonne, IL 60439, USA

[5] Department of Materials Science and Engineering, University of Illinois at Urbana-Champaign, Urbana, IL 61801, USA

## ABSTRACT

Direct detection of phonons is important for the investigation of information interconversion between the resonantly coupled magnons and phonons. Here we report resonant coupling of magnons and phonons, which can be directly visualized by using micro focused Brillouin light scattering in Ni/LiNbO$_3$ hybrid heterostructures. The patterns of surface acoustic wave phonons, originating from the interference between the original wave $\psi 0(A_0, \boldsymbol{k})$ and reflected wave $\psi 1(A_1, -\boldsymbol{k})$, can be modulated by magnetic field due to the magnon-phonon coupling. By analyzing the information of phonons obtained from Brillouin spectroscopy, the properties of the magnon system (Ni film), e.g., ferromagnetic resonance field and resonance linewidth can be determined. The results provide spatially resolved information about phonon manipulation and detection in a coupled magnon-phonon system.

---

[1] †liuqf@lzu.edu.cn

‡axelh@illinois.edu

# INTRODUCTION

Spin waves and their quasiparticles, i.e., magnons, are promising for high frequency information procession and transmission [1-6] and logic devices [1,7,8]. Unfortunately, the efficient generation of spin wave can be difficult and spin waves also propagate typically only over limited distances of <1 μm for most materials due to magnetic losses [9,10]. Travelling surface acoustic wave (SAW) phonons have the advantage of short wavelength (typically ~ 1 μm at 3 GHz) and long propagating distances of several mm, e.g., in $LiNbO_3$ crystals [11]. Coupling magnons to SAW phonons can avoid the rapid decay of spin wave, provides new opportunity for information interconversion of phononic and spin degrees of freedom [12-16], and may benefit spintronic device miniaturization and may broaden the scope of their applications [17-23]. However, current measurement method for magnon-phonon coupling based on magneto-transmission measurements cannot provide direct imaging of the magnetic modulation of the phonon systems [20-22,24,25]. Optical detection of hypersonic surface acoustic waves has been realized in bulk transparent materials [26] and transparent crystal substrate with reflection mirror [11,27,28] using Brillouin light scattering. This provides an opportunity to apply the Brillouin spectroscopy in a broader context for obtaining direct information of SAW, as well as the coupling of magnons and phonons.

In the present work, we focus on directly visualizing the resonant coupling of magnon and phonon using an optical approach. The interference patterns of multiple propagating SAW carrying spin information were detected using micro focused Brillouin light scattering (μ-BLS) on $Ni/LiNbO_3$ hybrid heterostructures. The BLS intensity has been found to be minimum at the ferromagnetic resonance (FMR) field of Ni, where SAW phonons are strongly attenuated by magnon. The interference patterns of the SAW almost disappeared near the FMR field of Ni, which provides a directly visualization of the magnetic field modulation of SAW phonons by resonant coupling of magnon-phonon. By fitting a detailed theoretical model [29-31] to the magnetic field dependence of the spatial-averaged and frequency-averaged BLS signal intensity, the estimated value of FMR field and resonance linewidth is consistent with that obtained based on SAW magneto-transmission measurements. The results provide new opportunities for phonon manipulation and detection in the presence of magnetization dynamics.

**RESULTS AND DISCUSSION**

Figure 1(a) shows an optical image of the Ni/LiNbO$_3$ hybrid device with a 50-nm thick rectangular Ni film. In these devices, the finger width and the spacing of the interdigital transducer (IDT) fabricated with 50-nm thick Al are both 3 μm (fundamental SAW wavelength $\lambda_0$ = 12 μm) on a 127.86º Y-X cut LiNbO$_3$ substrate with the thickness of 500 μm. By connecting to a microwave source using IDTs we excite SAW at 3.56 GHz, which corresponds to the 11$^{th}$ harmonic. Resonant coupling with the magnon system (Ni films) is established during the SAW propagation between the two IDTs. In order to achieve resonant coupling of magnons and phonons, a DC magnetic field $H$ is applied in a direction with an angle θ = 30º with respect to the SAW wavevector $k_{SAW}$. First, the laser spot was located on one of the IDT fingers, and the corresponding BLS signal is shown in Fig. 1(b). The BLS signal is independent of $H$ near 3.56 GHz, which indicates that SAW phonons are excited and detected, where the Al film acts as a mirror [11,27,28] for SAW phonons detection using μ-BLS. To depict the spatial distribution of the excited SAW phonons, a region on IDT fingers was selected to make spatially resolved measurements. Figure 1(c) shows an optical image of IDT fingers and the green rectangular region represents the area for the spatially resolved measurements. Figure 1(d) shows the image of surface acoustic wave phonons. The wavelength of 11$^{th}$ harmonic can be estimated as $\lambda_n$ = $\lambda_0$/n = 1.09 μm directly from the period of the spatially modulated BLS signal. The results indicate that spatial distribution of SAW phonons can be visualized using μ-BLS based imaging [11,26-28].

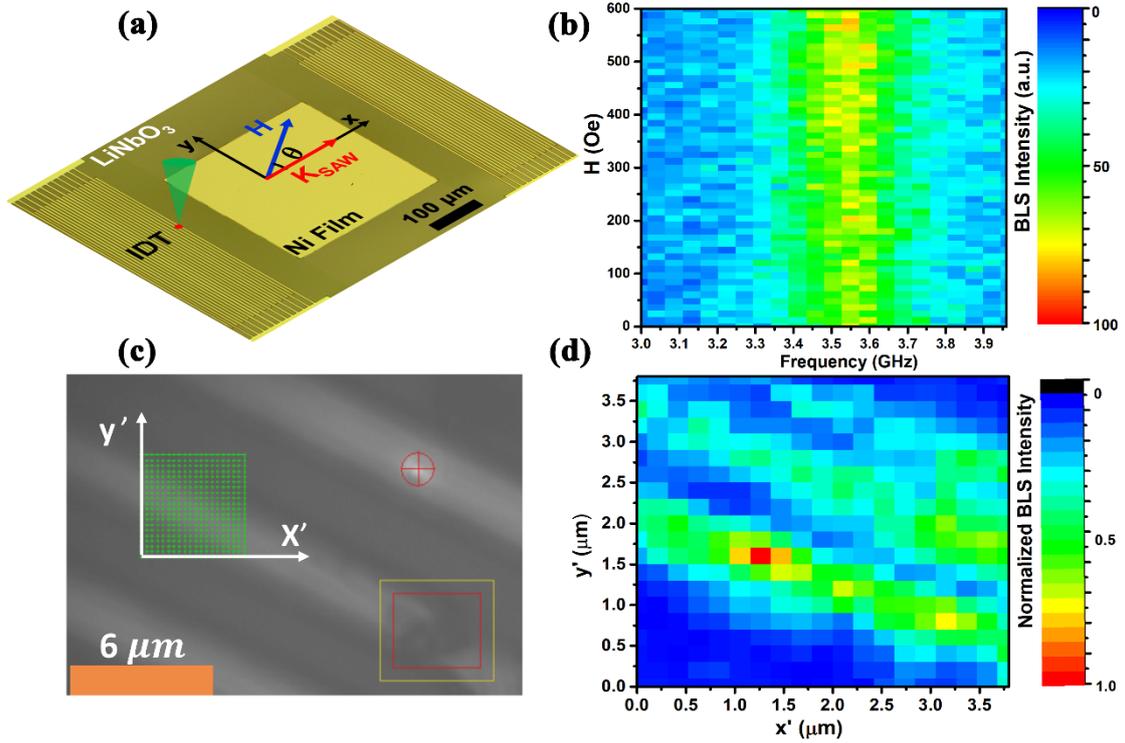

Figure 1: (a) Optical image of the Ni/LiNbO$_3$ hybrid device with a 50-nm thick Ni rectangular film. The green cone represents the BLS laser beam, and the red dot represents the position of laser spot. (b) Magnetic field dependence of the BLS signal when the laser spot is located on the IDT finger. (c) Optical image of IDT fingers; the green rectangular region represents the area for the of spatially resolved measurements. (d) Spatial resolved image of surface acoustic wave phonons.

Although the LiNbO$_3$ crystal is a perfect waveguide for SAW phonons propagation, it is difficult to directly detect the scattering of photons on the transparent substrate [26]. Figure 2(a) and (b) show the BLS intensity when the laser spot was located on the LiNbO$_3$ substrate and the Ni film, respectively. In order to exclude local variations, Figs. 2(a) and (b) are the averaged results for several measurement points (see the supplementary information) on the LiNbO$_3$ substrate and the Ni film, respectively. The BLS intensity on the LiNbO$_3$ is ten times weaker than it is on the Ni film at $H$ = 600 Oe. According to a Gaussian fit to the frequency distribution of the SAW phonons as shown in Figs. 1(c) and (d), the line width of the SAW phonons (FWHM = 0.28 GHz) at 3.56 GHz is consistent with that obtained on the Ni film (FWHM = 0.23 GHz). This indicates that the Ni film can be used as a mirror to detected SAW phonons on the transparent substrate [11,27,28]. Notably, the $H$ dependence of the BLS signal exhibits a gap near the FMR field of Ni [Fig 1 (b)]. The gap suggests that resonant coupling of magnons and phonons has been achieved, and that the SAW phonons are strongly attenuated by magnons around the FMR field. To quantify the phonon absorption efficiency due to FMR, the $H$ dependence of the frequency-averaged (3.15 GHz - 3.96 GHz) BLS intensity is plotted in Fig. 2(f). The obvious attenuation can be seen near $H$ = 200 Oe. The SAW phonons absorption efficiency due to resonant coupling can be estimated as 26.5 %. As a reference, the frequency-averaged BLS signal on LiNbO$_3$ is shown in Fig. 2(e), which shows no dependence on $H$.

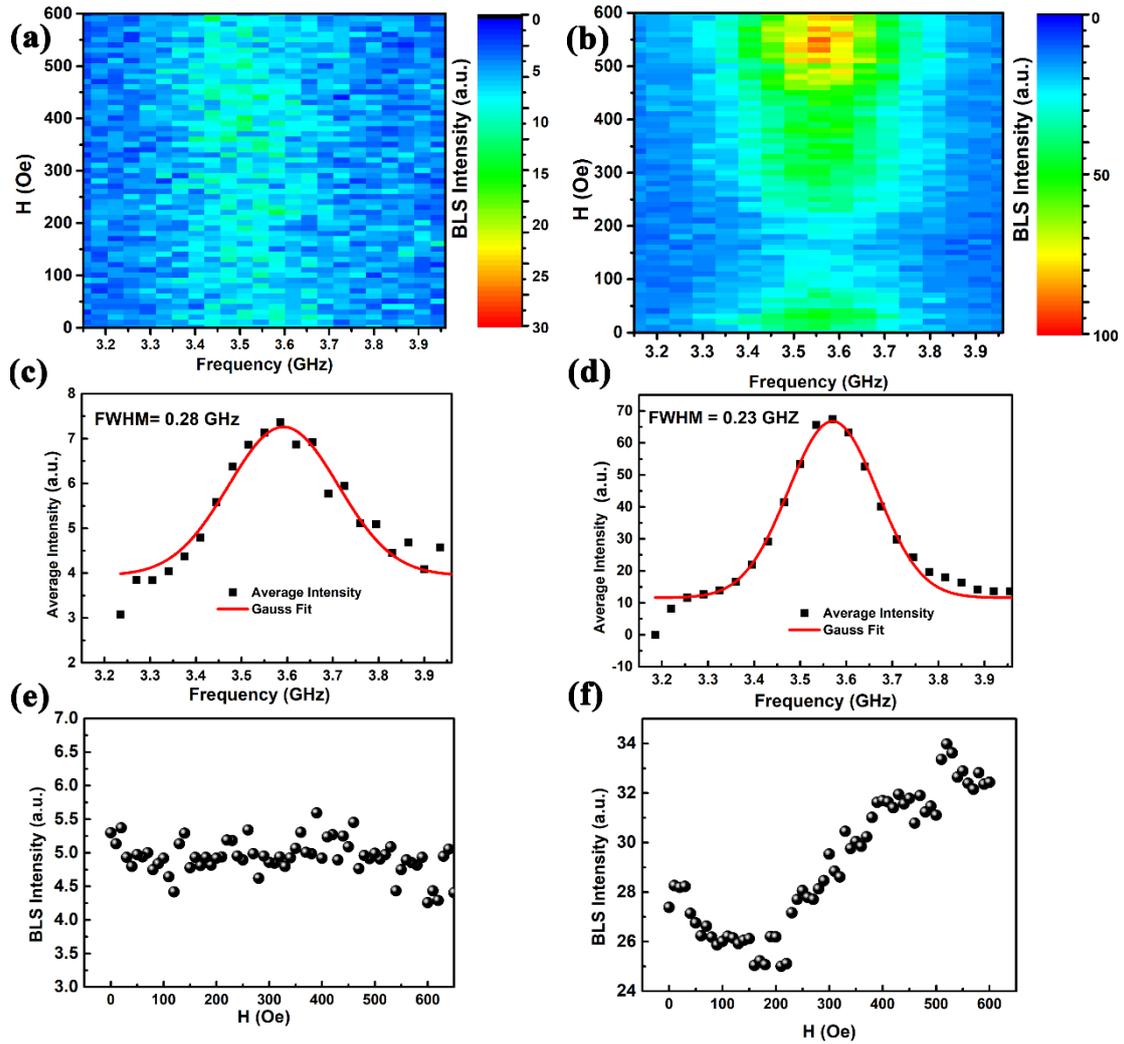

Figure 2: Magnetic field dependence of the BLS intensity when the laser spot is located (a) on the LiNbO$_3$ substrate and (b) on the Ni film, respectively; (c) and (d) Frequency distribution for (a) and (b), respectively; (e) and (f) Magnetic field dependence of the averaged BLS intensity for (a) and (b), respectively.

The magnetic field $H$ evolution for the interference patterns of SAW phonons are shown in Figs. 3(a-i). The period of the stripe pattern is ~ 1 μm, which is accordance with of the wavelength of 11$^{th}$ harmonic. The interference patterns of SAW almost disappeared near FMR field [Fig. 3(e) and (f)] of the Ni film. This is the direct evidence of phonon absorption by resonant magnons. In addition, the interference patterns become distorted below the saturation magnetic field [Fig. 3(h) and (i)], which can be attributed to the absorption of SAW phonons by hysteretic magnetization switching [20,32]. The results demonstrate how spatially resolved BLS imaging can visualize the resonant coupling of the SAW phonons and magnons.

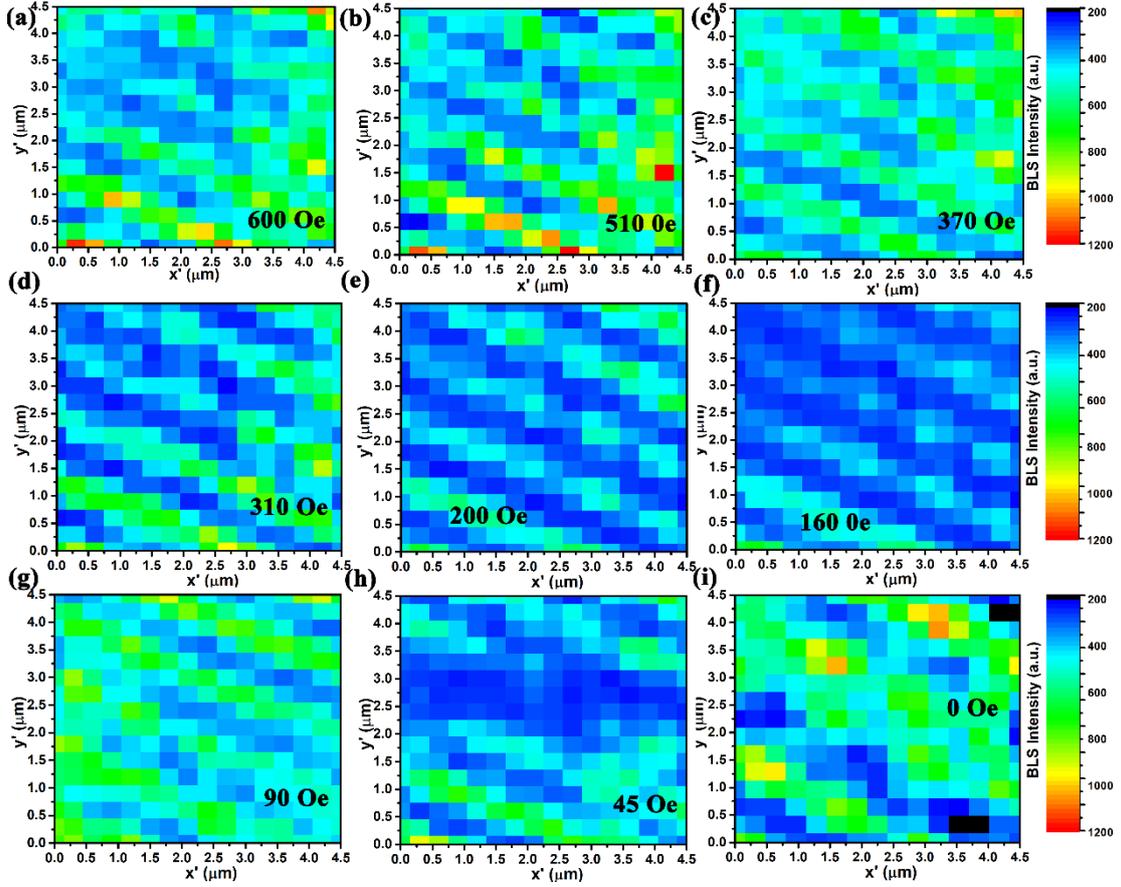

Figure 3: Magnetic field evolution of the BLS spatially resolved images when the laser spot is located on the Ni film: (a) 600 Oe; (b) 510 Oe; (c) 370 Oe; (d) 310 Oe; (e) 200 Oe; (f) 160 Oe; (g) 90 Oe; (h) 45 Oe; (i) 0 Oe.

In order to analyze the interference patterns of the SAW, the spatially resolved BLS intensity along the x′ direction (y′ = 1.5 μm) at different $H$ are plotted and fitted using a sinusoidal function:

$$I = I_0 + A \sin \frac{2\pi}{\lambda'(x'-x'_c)} \quad (1).$$

As shown in Fig. 4(a), the spatially resolved BLS intensity [Fig 3(e)] along the x′ direction (y′ = 1.5 μm) at $H$ = 200 Oe is plotted together with a fit to Eq. (1).

This suggests that the interference patterns cannot result from a standing wave, whose interference patterns should be described via: $I = |A\sin[2\pi/\lambda'(x'-x_c')]|$. According to the angle between the stripe pattern and the SAW propagating direction [x direction in Fig. 1 (a)], the SAW wavelength $\lambda_{SAW}$ and fitted $\lambda'$ can be depicted as: $\lambda_{SAW} = \lambda'\cos 60°$. Actually, $\lambda_{SAW} = \lambda_0/n = 1.09\ \mu m = (\lambda'\cos 60° \mp 0.03)\ \mu m$ is obtained from the fit results. Therefore, the patterns can come from the interference between the original wave $\psi 0(A_0, \mathbf{k})$ and the reflected wave $\psi 1(A_1, -\mathbf{k})$, where $A_0$ and $A_1$ are their amplitudes, and $\mathbf{k}$ is the wave vector. The minimum and maximum of BLS intensity of the interference patterns are proportional to $(A_0 - A_1)$ and $(A_0 + A_1)$, respectively. According to the fitting results, the values of $A_0$ and $A_1$ at different $H$ can be obtained. The ratio of $A_0/A_1$ is constant for different $H$, as is shown in Fig. 4(b).

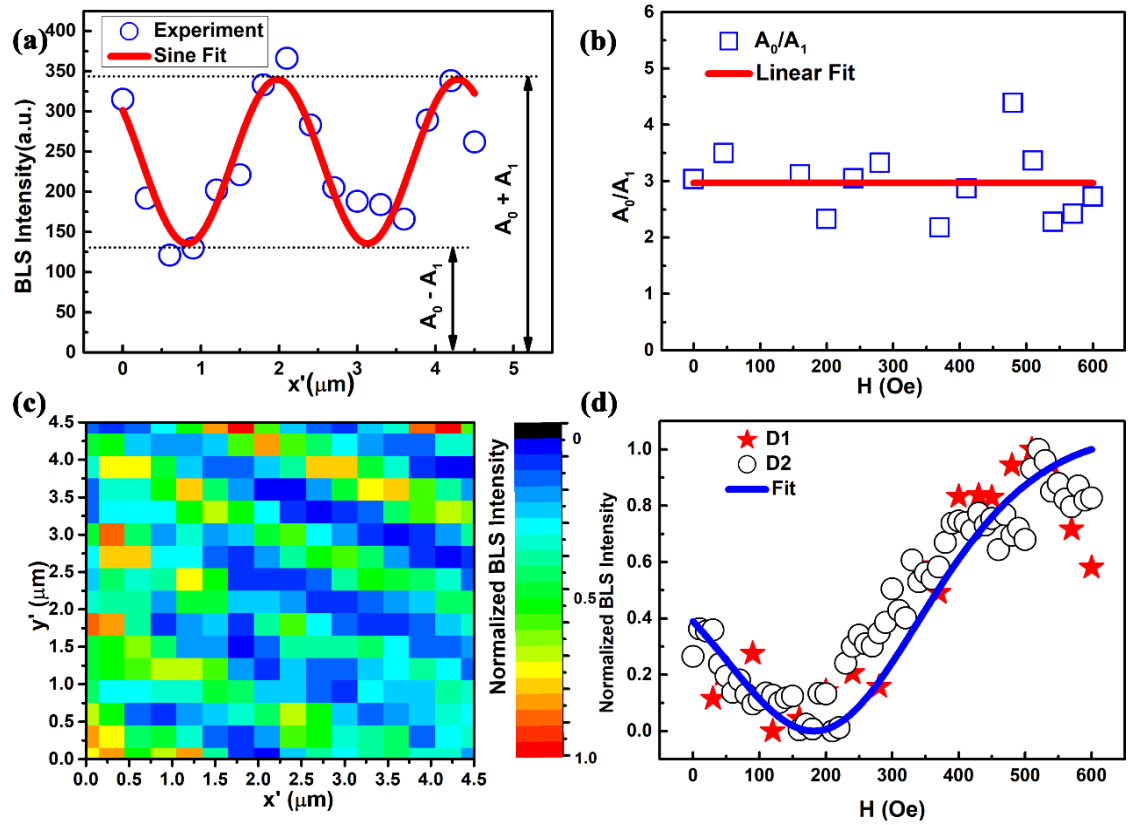

Figure 4: (a) BLS intensity evolution along the x' direction, and the solid line is a fit to Eq. (1). (b) $H$ dependence of $A_0/A_1$, and the solid line is linear fit. (c) Averaged spatial resolution image. (d) Field dependence of the normalized BLS intensity for the two measurements. The symbols are experimental data, and the solid line is a fit to theory using Eq. (2).

Since the $H$ evolution for the interference patterns of the SAW were measured at the same region, the stripe patterns have the same phase. When all the stripe patterns at different magnetic field are averaged, clear periodic patterns are obtained as shown in Fig. 4(c). This suggests that the patterns signal mainly come from SAW phonons. In order to fit the SAW phonon attenuation to $H$ in order to estimate values of the FMR field $H_r$ and the resonance linewidth $\Delta H$ of Ni, the spatially-averaged BLS signal intensity of each pixel in the spatial patterns at different $H$ [Fig. 3(a-i)] is calculated as data D1 [Stars in Fig 4(d)]. The magnetic field dependence of the frequency-averaged BLS intensity in Fig. 2(f) is normalized as data D2 [Circles in Fig. 4(d)]. The change of the SAW transmitted intensity (I) related to the phonon attenuation and generation during FMR can be rewritten as [29-31]:

$$I = I_0(1+p^2)^{-\frac{1}{2}}[1+(p+\beta)^2]^{-\frac{1}{2}} \exp\left[\frac{-\frac{\eta}{2}}{1+(p+\beta)^2}\right], \quad (2)$$

where, $p = (H - H_r)/\Delta H$, $\eta = C_1 f/\Delta H$, and $\beta = C_2 f^2/\Delta H$. And $I_0$, $C_1$, and $C_2$ are constant, $H_r$ is the FMR field, resonance linewidth $\Delta H$, $f = 3.56$ GHz is the frequency of the SAW, and $H$ is external magnetic field, respectively. The data D1 and D2 can be fitted well using Eq. (2), As shown in Fig. 4(d). The obtained $H_r$ and $\Delta H$ is $(236 \mp 30)$ Oe and $(290 \mp 36)$ Oe, respectively, which is consistent with the values ($H_r$= 270 Oe and $\Delta H$= 290 Oe) obtained from waveguide FMR experiments.

As shown here, the interference patterns of traveling multiple SAW coupled to magnon modes are well characterized using μ-BLS in the Ni/LiNbO$_3$ hybrid heterostructures. The Ni/LiNbO$_3$ hybrid heterostructures with resonant magnon-phonon coupling, can realize spatially complex transfers of energy between propagating spin and acoustic modes, thus creating a propagating magnetoelastic wave [15,23]. How to distinguish the spin signal and SAW signal from the BLS signal is an important issue for the future. Overall, our results open new directions for the application of SAW phonons in the fields of straintronics [16,18,20-22,24,25], sensing [33] and quantum information [34].

**CONCLUSION**

In conclusion, the interference patterns of traveling multiple surface acoustic waves coupled to magnon modes were detected using micro focused Brillouin light scattering in the Ni/LiNbO$_3$ hybrid heterostructures. The interference patterns of the surface acoustic wave almost disappear near the ferromagnetic resonance field of Ni, which provides a direct image of the magnetic field modulation of surface acoustic wave phonons by resonant magnon-phonon coupling. By fitting the theory to the magnetic field dependence of the averaged Brillouin light scattering intensity, the ferromagnetic resonance field and resonance linewidth were estimated to be consistent with those obtained based on surface acoustic wave magneto-transmission measurements. The patterns can be contributed to the interference between the original wave $\psi 0(A_0, \boldsymbol{k})$ and reflection wave $\psi 1(A_1, -\boldsymbol{k})$. These results provide a direct spatially resolved characterization of phonon manipulation and detection in the presence of magnetization dynamics.

**ACKNOWLEDGEMENTS**
This work was performed at the Argonne National Laboratory and supported by the Department of Energy, Office of Science, Materials Science and Engineering Division. The use of the Centre for Nanoscale Materials was supported by the US. Department of Energy (DOE), Office of Sciences, Basic Energy Sciences (BES), under Contract No. DE-AC02-06CH11357. Chenbo Zhao acknowledges additional financial support from the China Scholarship Council (no. 201806180105) for a research stay at Argonne. The authors thank Vasyl Tyberkevych and Guozhi Chai for useful discussion.

[1] A. V. Chumak, V. I. Vasyuchka, A. A. Serga, and B. Hillebrands, Nature Physics **11**, 453 (2015).
[2] D. Grundler, Nat Nanotechnol **11**, 407 (2016).
[3] Q. Wang, Pirro, P., Verba, R., Slavin, A., Hillebrands, B., & Chumak, A. V. , Science advances **4**, e1701517 (2018).
[4] A. V. Sadovnikov, A. A. Grachev, E. N. Beginin, S. E. Sheshukova, Y. P. Sharaevskii, and S. A. Nikitov, Physical Review Applied **7**, 014013 (2017).
[5] S. J. Hämäläinen, F. Brandl, K. J. A. Franke, D. Grundler, and S. van Dijken, Physical Review Applied **8**, 014020 (2017).
[6] Li, Y., Zhang, W., Tyberkevych, V., Kwok, W. K., Hoffmann, A., & Novosad, V. arXiv:2006.16158. (2020).
[7] K. Ganzhorn, S. Klingler, T. Wimmer, S. Geprägs, R. Gross, H. Huebl, and S. T. B. Goennenwein, Applied Physics Letters **109**, 022405 (2016).
[8] B. Rana and Y. Otani, Physical Review Applied **9**, 014033 (2018).
[9] A. Barra, A. Mal, G. Carman, and A. Sepulveda, Applied Physics Letters **110**, 072401 (2017).
[10] V. E. Demidov, S. Urazhdin, R. Liu, B. Divinskiy, A. Telegin, and S. O. Demokritov, Nat Commun **7**, 10446 (2016).
[11] B. Vincent, J. K. Krüger, O. Elmazria, L. Bouvot, J. Mainka, R. Sanctuary, D. Rouxel, and P. Alnot, Journal of Physics D: Applied Physics **38**, 2026 (2005).
[12] M. Foerster *et al.*, Nat Commun **8**, 407 (2017).
[13] A. Kamra, H. Keshtgar, P. Yan, and G. E. W. Bauer, Physical Review B **91**, 104409 (2015).
[14] M. Montagnese *et al.*, Phys Rev Lett **110**, 147206 (2013).
[15] P. Graczyk and M. Krawczyk, Physical Review B **96**, 024407 (2017).
[16] R. Verba, I. Lisenkov, I. Krivorotov, V. Tiberkevich, and A. Slavin, Physical Review Applied **9**, 064014 (2018)
[17] I. S. Camara, J. Y. Duquesne, A. Lemaître, C. Gourdon, and L. Thevenard, Physical Review Applied **11**, 014045 (2019).
[18] M. Weiler, H. Huebl, F. S. Goerg, F. D. Czeschka, R. Gross, and S. T. Goennenwein, Phys Rev Lett **108**, 176601 (2012).
[19] C. L. Chang, R. R. Tamming, T. J. Broomhall, J. Janusonis, P. W. Fry, R. I. Tobey, and T. J. Hayward, Physical Review Applied **10**, 034068 (2018).
[20] M. Weiler, L. Dreher, C. Heeg, H. Huebl, R. Gross, M. S. Brandt, and S. T. Goennenwein, Phys Rev Lett **106**, 117601 (2011).
[21] M. Xu, J. Puebla, F. Auvray, B. Rana, K. Kondou, and Y. Otani, Physical Review B **97** (2018).
[22] L. Dreher, M. Weiler, M. Pernpeintner, H. Huebl, R. Gross, M. S. Brandt, and S. T. B. Goennenwein, Physical Review B **86**, 134415 (2012)
[23] B. Casals, N. Statuto, M. Foerster, A. Hernandez-Minguez, R. Cichelero, P. Manshausen, A. Mandziak, L. Aballe, J. M. Hernandez, F. Macia, Phys Rev Lett **124**, 137202 (2020).
[24] R. Sasaki, Y. Nii, Y. Iguchi, and Y. Onose, Physical Review B **95**, 020407(R) (2017).
[25] P. G. Gowtham, T. Moriyama, D. C. Ralph, and R. A. Buhrman, Journal of Applied Physics **118**, 233910 (2015).
[26] R. J. Jiménez Riobóo, A. Sánchez-Sánchez, and C. Prieto, Physical Review B **94**, 014313 (2016).
[27] S. Murata, T. Kawamoto, M. Matsukawa, T. Yanagitani, and N. Ohtori, Japanese Journal of Applied Physics **46**, 4626 (2007).

[28] J. K. Krüger, B. Vincent, O. Elmazria, L. Bouvot, and P. Alnot, New Journal of Physics **6**, 57 (2004).

[29] I. A. Privorotskii, IEEE Transactions on Magnetics **16**, 3 (1980).

[30] I. A. Privorotskii, R. A. B. Devine, and G. C. Alexandrakis, Journal of Applied Physics **50**, 7732 (1979).

[31] C. Zhao, Y. Li, Z. Zhang, M. Vogel, J. E. Pearson, J. Wang, W. Zhang, V. Novosad, Q. Liu, A. Hoffmann, Physical Review Applied **13**, 054032 (2020).

[32] I. a. Feng, M. Tachiki, C. Krischer, and M. Levy, Journal of Applied Physics **53**, 177 (1982).

[33] V. P. B. Dominic Labanowski, Qiaochu Guo, Carola M. Purser, Brendan A. McCullian, P. Chris Hammel, Sayeef Salahuddin SCIENCE ADVANCES **4**, eaat6574 (2018).

[34] R. Manenti, A. F. Kockum, A. Patterson, T. Behrle, J. Rahamim, G. Tancredi, F. Nori, and P. J. Leek, Nat Commun **8**, 975 (2017).